\newcommand{\be}{\begin{eqnarray}}
\newcommand{\ee}{\end{eqnarray}}
\newcommand{\nn}{\nonumber}
\newcommand\q{\quad}

\documentclass[12pt]{article}

\usepackage{amsmath,amssymb}
\usepackage{graphicx}
\usepackage{cite}
\usepackage{bm}
\begin{document}
\renewcommand{\thefootnote}{\fnsymbol{footnote}}

\vskip 15mm

\begin{center}

{\Large Chiral anomalies in higher-derivative supersymmetric 6D gauge theories}

\vskip 4ex

A.V. \textsc{Smilga}

\vspace{.5cm}

SUBATECH, Universit\'e de
Nantes,  4 rue Alfred Kastler, BP 20722, Nantes  44307, France
\footnote{On leave of absence from ITEP, Moscow, Russia.}
 
$\texttt{smilga@subatech.in2p3.fr}$
\end{center}

\vskip 5ex

\begin{abstract}
We show that the recently constructed higher-derivative 6D SYM theory involves internal 
chiral anomaly
breaking gauge invariance. The anomaly is cancelled when adding 
to the theory an adjoint matter 
hypermultiplet.
 
\end{abstract}

\renewcommand{\thefootnote}{\arabic{footnote}}
\setcounter{footnote}0
\setcounter{page}{1}

\section{Introduction}
We argued recently \cite{TOE} that the fundamental Theory of Everything may represent a 
conventional field theory
defined in flat space-time with $D > 4$. Our Universe represents then a classical 3-brane solution in this
theory. Einstein's gravity appears as the effective lagrangian induced in the world-volume of  this brane. 
If this is true,
the fundamental higher-dimensional theory should be internally consistent. Two major problems which should be 
solved here are renormalizability and the existence and stability of quantum vacuum state (the absence of ghosts).

Conventional theories (involving the terms like Tr$\{F_{\mu\nu}^2\}$ etc) are not renormalizable for $D > 4$. 
However, if we add extra
derivatives such that the canonical dimension of the lagrangian is equal to $D$,  renormalizability might be 
achieved.
 Adding
extra derivatives creates the problem of ghosts \cite{PU}. However, one may hope that 
in {\it supersymmetric} theories, this problem
can be  handled. Indeed, supersymmetry algebra implies that the energies of all hamiltonian 
eigenstates are
 nonnegative.
Considering a particular supersymmetric QM 
model (dimensionally reduced  $5D$ superconformal Yang-Mills theory \cite{5DSC}) 

involving singularity and, potentially, associated ghost states 
\cite{5d}, we
showed that though the latter seem to be present in full spectrum  of the hamiltonian, the  
negative energy states
 do not possess normalizable superpartners and do not form complete supermultiplets. A reduced
Hilbert space involving only complete supermultiplets involves only the states with nonnegative energies. 
In Ref.\cite{5d}, 
we presented arguments that $5D$ superconformal YM theory  in the sector with zero v.e.v. of the
 real
scalar field $\sigma$ might be feasible and internally consistent. However, this theory is very diffucult to 
analyze 
because its
lagrangian does not involve quadratic terms and conventional perturbative methods do not work. In Ref.\cite{ISZ}, 
we constructed
using the methods of harmonic superspace \cite{harm } 
a scale invariant (conformally invariant at the classical level) supersymmetric YM theory in six dimensions.
 The component lagrangian has the form
\be
 {\cal L} &=& -   \, \mbox{Tr}
\left\{ \left( \nabla_\mu F_{\mu\lambda}\right)^2  +  i\psi^j\gamma_\mu \nabla_\mu (\nabla)^2\psi_j
 + \frac 12 \left(\nabla_\mu{ D}_{jk}\right)^2
\right. \nn \\
&& \q \left.
+\,  { D}_{lk}{D}^{kj}{D}^{\;\;\;l}_{j}
  -2i {D}_{jk} \left( \psi^j\gamma_\mu\nabla_\mu \psi^k
- \nabla_\mu \psi^j \gamma_\mu \psi^k \right)
+ (\psi^j\gamma_\mu \psi_j)^2  \right. \nn \\
&& \q \left. +\, \frac 12  \nabla_\mu\psi^j
\gamma_\mu\sigma_{\nu\sigma}[F_{\nu\sigma}, \psi_j]
- 2\nabla_\mu F_{\mu\nu}\, \psi^j\gamma_\nu\psi_j
 \right\}
\label{CompAct}
 \ee
 where  $\sigma_{\mu\nu} = (\tilde \gamma_\mu \gamma_\nu -  \tilde \gamma_\nu \gamma_\mu)/2 $. 
The lagrangian (\ref{CompAct}) involves gauge fields $A_\mu$, adjoint Weyl fermions $\psi^{ka}$, where $a = 1,2,3,4$ is 
 the spinor index  and the subflavor index $k$ acquires two values,
and adjoint scalar fields $D_{jk} = D_{kj}$ of canonical dimension 2. (In the conventional SYM theory
the fields $D_{jk}$ are auxiliary and enter without derivatives. But in the higher-derivative theory 
(\ref{CompAct}), they  propagate.) Here all the fields  are Hermitian $N\times N$ colour matrices. 
The $6D$ Weyl  $4\times 4$ matrices $(\gamma_\mu)_{ab}$ are antisymmetric. They satisfy the relation
 \be
\label{antigam}
\gamma_\mu \tilde\gamma_\nu +  \gamma_\nu \tilde\gamma_\mu = - 2 \eta_{\mu\nu}\ ,
\ee
where 
$$ \tilde\gamma_\mu^{ab}=\frac 12 \varepsilon^{abcd}
(\gamma_\mu)_{cd}\ . $$
One of the possible explicit representations is
 \be
\label{gammy} 
\gamma_0 = \tilde \gamma_0 = i \sigma_2 \otimes 1\!\!1 \ ; \ \ \ \ \ 
\gamma_1 = -\tilde \gamma_1 = i \sigma_2 \otimes \sigma_1 \ ; \ \ \ \ \ 
\gamma_2 = -\tilde \gamma_2 = i  1\!\!1 \otimes \sigma_2 \ ; \nonumber \\
 \gamma_3 = -\tilde \gamma_3 = i \sigma_2 \otimes \sigma_1 \ ; \ \ \ \ \ 
\gamma_4 = \tilde \gamma_4 =  \sigma_1 \otimes \sigma_2 \ ;\ \ \ \ \ 
\gamma_5 = \tilde \gamma_5 = \sigma_3 \otimes \sigma_2 \ .
 \ee
The property 
 \be
\label{Trgameps}
\frac 14 {\rm Tr} \{ \gamma_\mu \tilde \gamma_\nu \gamma_\alpha \tilde \gamma_\beta 
\gamma_\gamma \tilde \gamma_\delta 
\} \ = - \frac 14 {\rm Tr} \{ \tilde \gamma_\mu  \gamma_\nu \tilde \gamma_\alpha  \gamma_\beta 
\tilde \gamma_\gamma  \gamma_\delta 
\}\ =\ \epsilon_{\mu\nu\alpha\beta\gamma\delta}\ + {\rm symmetric\ part}
 \ee
holds (with the convention $\epsilon_{012345} = 1$).

The fermions  $\psi^{ja}$ belong to the representation (0,1) of the group $SO(5,1)$. They satisfy the pseudoreality
constraint
 \be
\label{pseudo}
\left(\psi^{aj}\right)^*  =  \epsilon_{jk}   (\gamma_0)_{ab} \psi^{bk} \ .
 \ee
Note that the possibility to impose such a constraint is specific for six dimensions. Indeed, both
$SO(3,1)$ and $SO(5,1)$ involve two different chiral spinor representations. But their behavior under complex
conjugation is different for $D=4$ and $D = 6$. When $D= 4$, 
complex conjugation changes the type of representation of a Minkowski spinor $(1, 0) \leftrightarrow (0, 1)$. 
On the other hand, for $D=6$, the complex conjugated spinor belongs to the same representation as the original one. One can say that the
subflavor index $j =1,2$ corresponds to the original spinor and the complex conjugated one.
\footnote{
Note, that in Euclidean
space the situation is exactly inverse. $SO(4) \equiv SU(2) \otimes SU(2)$  with two completely independent factors.
 The spinor represents a doublet
under either left or right such $SU(2)$ factor and complex conjugation does not change this. On the other hand,
$SO(6) \equiv SU(4)$ and complex conjugation transforms ``quarks'' into ``antiquarks''. }

The lagrangian (\ref{CompAct}) involves a dimensionless coupling constant, which suggests renormalizability.
In \cite{ISZ}, we calculated the beta function in this theory and found that it has the same sign as in the ordinary
$4D$ QED corresponding to the zero charge situation. However (and this is the main observation of this paper), 
the theory (\ref{CompAct}) is {\it not} renormalizable
because it involves chiral anomaly ! 
 \footnote{The original guess belongs  to Soo-Jong Rey \cite{Rey}.}
The reason is the mentioned above chiral nature of Minkowskian fermions in six dimensions, irrespectively whether
they belong to a real or to a complex representation of the gauge group. Indeed, the lagrangian (\ref{CompAct})
 involves
only {\it one} type of spinors $\psi^a$ while the spinors $\chi_a$ belonging to the representation (1,0) 
are absent. 
It is this  asymmetry 
which gives rise to anomaly (cf. the  four-dimensional  situation, where the SYM lagrangian  involves
 the structure $\sim {\rm Tr} \{ \psi \sigma_\mu \nabla_\mu  \bar\psi \}$ containing both $\psi$ and $\bar \psi$;
 there is no asymmetry and no chiral anomaly). 
    
\section{Anomalies.}

The anomalies in standard  higher-dimensional theories involving fermion kinetic term
$\propto \bar \psi D \!\!\!\!/ (1 + \Gamma^{D+1}) \psi$ were 
thouroughly studied in  \cite{Framp} - \cite{Zumino}. In six
dimensions, one has to calculate the anomalous box graph. Another convenient method uses the Schwinger 
splitting technique. Consider the theory with the standard fermion kinetic term
  \be
\label{Lferm2}
{\cal L} \ =\  i {\rm Tr} \, \{  \psi^{k} D \!\!\!\!/ \psi_k \}\ , 
 \ee 
 where  $D \!\!\!\!/ = \gamma_\mu \nabla_\mu$,  $\nabla_\mu  = \partial_\mu - i  A_\mu$. The colour current
$J_\mu^A  = {\rm Tr} \{\psi^k T^A \gamma_\mu \psi_k \}$ ($T^A$ is the adjoint generator)
 is covariantly conserved at the classical level. Quantum effects bring about the anomaly:
   \be
 \label{6Danom}
\nabla_\mu J_\mu^{A} \ = \  \frac 1{3\cdot 128 \pi^3} \epsilon_{\mu\nu\alpha\beta\gamma\delta} 
{\rm Tr} \{T^A   F_{\mu\nu}   F_{\alpha\beta}   F_{\gamma\delta} \}\ + \ldots
 \ee
where the dots stand for the terms of higher order in $A_\mu$ having the form   \\ $\sim 
\epsilon_{\mu\nu\alpha\beta\gamma\delta}{\rm Tr}
\{ T^a    F_{\mu\nu}   F_{\alpha\beta}   A_\gamma    A_\delta \}$, 
$\sim \epsilon_{\mu\nu\alpha\beta\gamma\delta}{\rm Tr}
\{ T^a    F_{\mu\nu}    A_\alpha    A_\beta   A_\gamma    A_\delta \}$ and \\ 
$\sim \epsilon_{\mu\nu\alpha\beta\gamma\delta}{\rm Tr}
\{ T^a     A_\mu    A_\nu    A_\alpha    A_\beta   A_\gamma    A_\delta \}$. 
The coefficients of these terms are rigidly
related \cite{Zumino} to the coefficient in Eq.(\ref{6Danom}).

Note that the anomaly is proportional to the symmetrized trace Tr$\{T^{(A}T^B T^C T^{D)} \}$ which 
does not vanish. If we would try to evaluate the internal chiral anomaly in the $4D$ SYM theory, it would
involve the factor Tr$\{T^{(A} T^B T^{C)} \} \ =\ 0$. In other words, the anomaly  vanishes there and
this is of course related to the fact that in four dimensions the Minkowski 
 fermion kinetic term involves the fermion fields both
in $(0,1)$ and $(1, 0)$ representations, as explained above.
 Note also  that, for the fermions belonging to the representation $(1,0)$ of $SO(5,1)$, the result
would be the same with the opposite  sign, as follows from (\ref{Trgameps}). 

In the case under consideration, the lagrangian is different from (\ref{Lferm2}) and involves higher derivatives.
However, as was observed in \cite{AGW}, this does not change the result for the anomaly. The argument of 
Ref.\cite{AGW} 
was the following. Consider the $U(1)$ theory (non-Abelian structures change nothing) 
with the lagrangian, involving both left and right handed fermions 
with different
kinetic terms. In Dirac notations,
 \be
\label{LAB}
{\cal L} \ =\ \bar \psi A (1 + \Gamma^{D+1}) \psi + \bar \psi B (1- \Gamma^{D+1}) \psi \ =
\bar \psi (A+B) \psi + \bar \psi (A-B) \Gamma^{D+1} \psi
 \ee
For example, one can take $A = D \!\!\!\!/$ and   $B = D \!\!\!\!/^3$. We are allowed to regularize the theory
in the ultraviolet multiplying $A+B$ in the first term by $(1 - {\cal D}_\mu^2/\Lambda^2)^n$ while keeping the second 
term intact. This will bring in the factor $(1 + p^2/\Lambda^2)^n$ in  the fermion propagator while the axial current
depending only on the second term would not change. Then the Feynman integral for the anomalous box graph
would involve four such factors downstairs and not more than three such factors upstairs coming from the vector
vertices. As a result, the integral would converge in the ultraviolet meaning  the absence of anomaly. And 
this means that the anomaly coefficients in the theories $\bar \psi A(1 + \Gamma^{D+1}) \psi$ and   
$\bar \psi B(1 + \Gamma^{D+1}) \psi $ are the same.

Another argument is the following \cite{Rey}. Consider a mixed theory  
 \be
\label{mixed}
{\cal L}_{\rm ferm} \ =\ \frac i2 \bar \psi (D \!\!\!\!/ - D \!\!\!\!/^3/M^2)(1 + \Gamma^{D+1}) \psi\ . 
  \ee
 When ultraviolet regulator is large $\Lambda \gg M$,
the higher-derivative term dominates in all calculations and the anomaly should be the same as in the pure 
higher-derivative theory. On the other hand, the anomaly has an infrared face and is related to the number of 
levels crossing zero in a slowly varying external field with different Chern-Simons numbers at $t = - \infty$ 
and
$t = \infty$ . But when the characteristic frequency of this external field is much smaller than $M$, 
one can forget
about the higher-derivative term in (\ref{mixed}) and the anomaly should be the same as in the theory with
 the standard Dirac
kinetic term. Still another way to derive the same result is noting that the  indices
 of the Euclidean operators $D \!\!\!\!/$, 
$D \!\!\!\!/^3$, $D \!\!\!\!/ D_\mu^2$, etc coincide. 

Finally, one can amuse  oneself by the explicit calculation of the 
anomalous divergence of the gauge current for the lagrangian (\ref{mixed}) using 
the Schwinger splitting technique and
be convinced that the coefficient is, indeed,  the same as for the standard Dirac lagrangian (see Appendix).

\section{Discussion}

The presence of internal chiral anomaly in the theory means breaking of gauge invariance. Among other things, this makes
the theory  nonrenormalizable.
Though there were attempts to attribute meaning to anomalous theories \cite{Faddeev}, we prefer to stick
to a conservative viewpoint by which such theories are sick.  
 In a ``healthy'' theory, chiral anomaly must be cancelled. To achieve such cancellation, one has to include in the theory
some other fermions in addition to the gluino fields $\psi^{ka}$, the  superpartners of the gauge potential. In order to keep
supersymmetry, these extra fermions should come together with their superpartners. The only  $6D$ supermultiplet besides
the vector multiplet which admits off-shell formulation is the hypermultiplet. We considered the problem of coupling
$6D$ hypermultiplet to $6D$ vector multiplet in Ref.\cite{hyphyp}. We found out there that it is difficult to do it
in a ``symmetric'' way such that the kinetic term  involves higher derivatives for all  fermions present in the theory.
 The problem is that an off-shell hypermultiplet involves besides physical fields 
an infinite number of auxiliary
components (that is especially clearly seen in the harmonic superspace approach \cite{harm}). For a 
conventional hypermultiplet, these extra components are auxiliary, indeed. They 
enter in the lagrangian without derivatives and can be easily integrated over. However, in a HD lagrangian, the former 
auxiliary fields acquire derivatives and become  propagating.

It is not quite clear for us what does a theory with an infinite number of propagating
massless degrees of freedom mean. For example, an infinite number of propagating fields may
provide an infinite contribution to  the beta function ( though it is  not known at present 
whether it is the case or not).
 It would be interesting to think more in this direction. What we would like to point out now, however, 
is that an adjoint 
hypermultiplet
gives a {\it finite} contribution to the anomaly, which exactly cancels the pure SYM contribution (\ref{6Danom}).

As was discussed in \cite{hyphyp}, there are many ways to couple the hypermultiplet to the gauge supermultiplet.
\footnote{Incidentally, none of them allows one to preserve the 
conformal invariance of the classical action (though scale invariance is manifest).}
 Consider 
e.g. the action $S_2$ given by Eqs.(B3,B6) of Ref.\cite{hyphyp}. The fermion kinetic term is
 \be
\label{S2kin}
 {\cal L}_{\rm kin} \ \sim \ \int du {\rm Tr} \{ i\chi^A_a  ({\tilde \gamma}^\mu)^{ab}\partial_\mu \Box 
 \chi_{Ab} -   
2 \chi^A_a \partial^{++} \Box \lambda_A^a \}\ ,
 \ee 
where $A = 1,2$ is the subflavor  index. The fields $\chi^A, \lambda_A$, subject to pseudoreality constraints like in 
Eq.(\ref{pseudo}), 
are expanded over the harmonics $u^{\pm}_i$, $i = 1,2$
 \footnote{ the harmonics satisfy the relation $u_i^- u^{+i} = 1$ and represent the complex coordinates on 
$C\!\!\!\!C P\!\!\!\!P^1$ - the coset of the automorphism group of the $\mathcal{N} = 1$ $6D$ SUSY algebra.} 
as
 \be
\label{psichiexp}
 \chi(u,x) &=& \chi(x) + \chi^{(ij)}(x) u_i^+ u_j^- + \chi^{(ijkl)}    u_i^+ u_j^+  u_k^- u_k^- + \ldots \nonumber \\
 \lambda(u,x) &=& \lambda^{(ij)}(x) u_i^- u_j^- + \lambda^{(ijkl)}  u_i^- u_j^-  u_k^- u_l^+ + \ldots
   \ee
Note that the field $\chi$ has the spinor indices down and 
belongs to {\it another} fermion representation compared to the gluino field $\psi^{ka}$. Indeed, the field $\chi_a^k$ 
comes from the linear term of the expansion of the hypermultiplet superfield in $\theta^a$, while the
field $\psi^{ka}$ comes from the cubic in $\theta$ term $\sim \varepsilon_{abcd} \psi^a \theta^b \theta^c \theta^d$
\cite{ISZ,hyphyp}. The field $\lambda^a$ stems from the cubic in $\theta$ term of the hypermultiplet expansion 
and has the same chirality
as the gluino field.
 
In each term of the expansion of $\chi$, the number of the factors $u^-$ is equal to the number of the factors $u^+$.
(the eigenvalue of the  harmonic charge $u^+_i \partial/\partial u^+_i 
- u^-_i \partial/\partial u^-_i$ for the field  $\chi$ is zero). It exceeds by 2  the number of the factors $u^+$ for 
the expansion of $\lambda$ (the latter has harmonic charge -2). The harmonic derivative $\partial^{++}$ entering 
(\ref{S2kin})
is defined as $u^+_i \partial/\partial u^-_i$. Harmonic integrals $\int du$ are nonzero only for the structures 
of zero harmonic charge,
 \be
 \label{intu}
\int du \, = 1,\ \ \ \ \ \int du \, u^+_i u^-_j = \frac 12 \epsilon_{ij}, \ \ \ \ \ 
\int du \, u^+_i u^+_k u^-_j u^-_l = \frac 16 (\epsilon_{ij} \epsilon_{kl}
+ \epsilon_{il} \epsilon_{kj} ), \ \ \ldots
 \ee
 Thus, we have an infinite number of physical 
fields $\chi(x), \chi^{(ij)}(x), \lambda^{(ij)}(x)$, etc. with growing isospins. 
When substituting the expansion (\ref{psichiexp}) into
Eq.(\ref{S2kin}), we obtain 
 \be
\label{kinexpan}
{\cal L}_{\rm kin} \ =\ i \chi^A \tilde \gamma^\mu \partial_\mu \Box \chi_A - \frac i3 \chi^A_{(ij})  
\tilde \gamma^\mu \partial_\mu \Box
\chi_A^{(ij)} + \frac 43 \chi^A_{(ij}) \Box \lambda_A^{(ij)}
 \ee
The first term gives a nonzero contribution to the anomaly, which cancels the contribution in 
Eq.(\ref{6Danom}).
On the other hand, the contribution of the terms involving $\chi^{(ij)}$ and $\lambda^{(ij)}$ vanishes because the fields 
$\chi^{(ij)}$ and $\lambda^{(ij)}$ have opposite chiralities and we argued above that the anomaly does not depend on a
 particular form
of the lagrangian, but only on the field content. (To illustrate this, one can e.g. lift the box operator 
in the second and the third terms
in Eq.(\ref{kinexpan}). Then variation over $\lambda^{(ij)}$ gives the equation of motion $\chi_{(ij)} = 0$ so that the
 current and
its anomalous divergence vanish.) The same is true for the terms involving the components $\chi^{(ijkl)}$ and
 $ \lambda^{(ijkl)}$, etc.

Another ``asymmetric'' possibility is to couple the HD vector multiplet  to a conventional 
hypermultiplet  with the standard kinetic term. In this case, one can get
rid of the auxiliary fields, as usual. We are left with only one 
 Weyl fermion $\chi^A_a$ satisfying the pseudoreality constraint. Its contribution to the anomalous divergence
cancels the gluino contribution.
 The conformal invariance of the classical lagrangian
can be imposed if attributing canonical dimension 2 to scalar components and dimension $5/2$
to  $\chi^A_a$. 
Unfortunately, classical conformal symmetry is not preserved at the quantum level. Conformal
anomaly (alias, beta function) was calculated in \cite{ISZ,hyphyp}. We found that the contributions to the beta function
coming from interactions with hypermultiplet and from vector multiplet self-interactions have the same sign
corresponding to the Landau zero situation, like in the ordinary QED.   
   
 Besides logarithmic renormalization of the coupling constant (the coefficient at the structure (\ref{CompAct})
in the effective lagrangian), the theory may involve also quadratic ultraviolet divergences at the     
 structure $\sim {\rm Tr} \{F_{\mu\nu}^2 \} + \ldots$ (the  conventional  $6D$ SYM lagrangian). We found earlier that
this coefficient vanishes for the pure HD SYM theory (\ref{CompAct}), but does not vanish for the theories
involving hypermultiplet interactions. In other words, the ``asymmetric'' anomaly-free theory involves a quadratic divergence,
which should be cancelled by a properly chosen counterterm. 
This  theory has thus roughly the same status as the scalar
QED or $\lambda \phi^4$ theory in four dimensions, where renormalization of the scalar mass involves quadratic UV
divergences.  All these theories are renormalizable, but the eventual cancellation
 of the divergences implies the presense of the counterterms
 $\sim \Lambda_{UV}^2$ with a fine-tuned coefficient. In all these
theories, the effective charge grows at high energies, which means that, in spite of being renormalizable, 
the theories are not
consistently defined nonperturbatively. 

Constructing a nontrivial $6D$ theory that would be internally consistent both  perturbatively
and nonperturbatively remains a major challenge. 

I am indebted to  E. Ivanov and S. Theisen for useful discussions and to Soo-Jong Rey for many illuminating discussions
and comments. I acknowledge warm hospitality at AEI institute at Golm where this work was finished.

\section*{Appendix}
We will demonstrate here the independence of the anomaly coefficient of the particular form of the lagrangian
by illustrative calculations using Schwinger-splitting UV regularization  and background field technique 
\cite{Peskin}.
Let us be interested in the global axial anomaly in the $U(1)$ vector theory
with the lagrangian 
  \be
\label{mixvect}
{\cal L}_{\rm ferm} \ =\ i\bar \psi (D \!\!\!\!/  - D \!\!\!\!/^3/M^2) \psi\ . 
  \ee
 The corresponding equations of motion are 
 \be
\label{eqmot}
\left( D \!\!\!\!/ - {D \!\!\!\!/^3}/{M^2} \right) \psi = 0\ ; \ \ \ \ \ \ \ \ 
\bar \psi \left( \stackrel \leftarrow {D \!\!\!\!/} -  {\stackrel \leftarrow {D \!\!\!\!/}^3}/{M^2} 
\right ) = 0 \ . 
   \ee
  The N\"other axial current is
  \be
\label{current}
 {\cal A}_\mu \ =\ \bar \psi \Gamma_\mu \Gamma^{D+1} \psi - \frac 1 {M^2} \bar \psi 
\left[ \Gamma_\mu D \!\!\!\!/^2
- \stackrel \leftarrow {D \!\!\!\!/} \Gamma_\mu  {D \!\!\!\!/} \ + \stackrel \leftarrow {D \!\!\!\!/}^2 
\Gamma_\mu \right]
 \Gamma^{D+1}  \psi \ ,
 \ee
 where $\Gamma_\mu$ are Dirac gamma matrices in $D$ dimensions (not to confuse with 6-dimensional 
Weyl matrices $\gamma_\mu$ ). It is conserved at the classical level.
The Schwinger-splitted current is 
 \be
\label{cureps}
 {\cal A}_\mu^{(\epsilon)}  \ =\ E \bar \psi_+ \Gamma_\mu \Gamma^{D+1} \psi - 
\frac E {M^2} \bar \psi_+ 
\left[ \Gamma_\mu D \!\!\!\!/^2
- \stackrel \leftarrow {D \!\!\!\!/}_+ \Gamma_\mu  {D \!\!\!\!/}\  + \stackrel \leftarrow {D \!\!\!\!/}^2_+ 
\Gamma_\mu \right]
 \Gamma^{D+1}  \psi \ ,
 \ee
where $\psi_+$ and $ \stackrel \leftarrow {D \!\!\!\!/}_+$ are evaluated at the point $x+\epsilon$ and
 \be
\label{E}
 E \ =\ P \exp \left\{ i \int_x^{x+\epsilon} A_\alpha (y) dy_\alpha \right \}\ .
 \ee
is the path-ordered exponent. It is convenient to work 
 in the Fock-Schwinger or fixed point gauge $x_\mu A_\mu = 0$ \cite{fixed,fgja} where the vector
potential is expressed via the field density   as 
 \be
\label{fpoint}
A_\mu(x) = \frac 12  F_{\nu\mu} x_\nu + ...\ ,
  \ee
 the dots standing for the terms involving $\partial_\alpha F_{\mu\nu}, \partial_\alpha 
\partial_\beta F_{\mu\nu}$, etc. We can safely  neglect them 
because
 the sought-for anomaly  involves only $F$, but not its  derivatives. 
The anomalous divergence is 
 \be
\label{andir}
 \partial_\mu {\cal A}_\mu \ =\ \lim_{\epsilon \to 0}\,  i F_{\mu\nu} \epsilon_\nu   
\left\{\Gamma_\mu  - \frac 1{M^2} \left[ \Gamma_\mu D \!\!\!\!/^2
- \stackrel \leftarrow {D \!\!\!\!/}_+ \Gamma_\mu  {D \!\!\!\!/} \ + \stackrel \leftarrow {D \!\!\!\!/}^2_+ 
\Gamma_\mu \right] \Gamma^{D+1} 
{\mathcal G} (x,x+\epsilon) \right \}\ ,
  \ee
where ${\mathcal G}(x, x+\epsilon)$ is fermion Green's function 
$\langle \psi(x) \, \bar \psi (x+\epsilon) \rangle_A$ evaluated in the presence of  the 
background  $A_\mu$. When deriving (\ref{andir}), we took into account the terms where 
the derivatives act on the fermion fields
(and used the equations of motion (\ref{eqmot})) and also the terms coming from differentiating
the factor 
$$E \approx 1 + \frac i2  F_{\nu\mu} x_\nu \epsilon_\mu \ .$$
 One can show that  these two  contributions are equal.

The calculations are trivial in two dimensions where the anomaly is linear in $F$. 
As the factor $\sim F$ is already present in (\ref{andir}), 
we can trade  the covariant derivatives there for the usual ones and substitute  tree Green's function
 \be
\label{Gtree}
 G_0(-\epsilon) = \int \frac {d^2p}{(2\pi)^2} G_0(p) e^{ip\epsilon}\ ,\ \ \ \ \ \ \ \ \ G_0(p) 
= \frac {ip\!\!\!/ }{p^2(1 + p^2/M^2)} 
  \ee
for ${\mathcal G}(x, x+\epsilon)$. Using $\stackrel \leftarrow {\partial \!\!\!/} \! \Gamma_\mu \equiv 
2 \partial_\mu - \Gamma_\mu  \!\!\stackrel \leftarrow {\partial  \!\!\!/}$, antisymmetry of $F_{\mu\nu}$,
 and going
into momentum space, one immediately sees that the anomaly (\ref{andir}) involves the factor 
$\sim 1 + p^2/M^2$ upstairs, which cancels the same factor downstairs in Eq. (\ref{Gtree}). In other words,
the result does not depend on $M$.

The case $D=4$ is slightly less trivial. One has to take into account two contributions: 
{\it (i)} the terms where covariant derivatives in Eq.(\ref{andir}) are traded for 
the usual ones, but Green's function is 
evaluated in linear order in $F$ ; {\it (ii)} the terms with tree level Green's function (\ref{Gtree}) where 
the extra power of $F$ is extracted from 
    \be
  \label{DFSig}
   {D\!\!\!\!/}^2 = \partial^2 - \frac i2  {F_{\alpha\beta} \Gamma_{\alpha} \Gamma_{\beta }}\,,\ \ \ \ \ \ \ 
   \stackrel \leftarrow {D\!\!\!\!/}^2 = \partial^2 - \frac i2  {F_{\alpha\beta} \Gamma_{\alpha} \Gamma_{\beta }} \ .
   \ee
The contribution of the  second type is convenient to present as
  \be
 \label{ii}
\left. \partial_\mu {\cal A}_\mu \right|_{G_0} = 
- \frac {F_{\mu\nu} \epsilon_\nu}{M^2} \int \frac {d^4p}{(2\pi)^4} e^{ip\epsilon} \, 
{\rm Tr} \{\Gamma_\mu 
F_{\alpha\beta} \Gamma_{\alpha} \Gamma_{\beta } \Gamma^5 G_0(p) \}\ .
 \ee
To fix the contribution of the first type, we have to evaluate Green's function in the first order in $F$.
It is given by the graph  in Fig. 1. We have 
  \be
 \label{G1}
{\mathcal G}_1(x, x+\epsilon)  \ =\  i\int d^4u \, G_0(-u)  \Lambda_\alpha A_\alpha(u+x) G_0(u-\epsilon)
 \ =\  \nn \\
- \frac i2 \int d^4u \, G_0(-u)  \Lambda_\alpha  F_{\alpha\beta} (u+x)_\beta G_0(u-\epsilon)\ ,
   \ee 
where the vector potential $A_\alpha(x)$ is chosen in the fixed point gauge form (\ref{fpoint}) and
$\Lambda_\alpha$ is the vertex following from the lagrangian (\ref{mixvect}). In the momentum representation,
 \be
\label{Lambda}
 \Lambda_\alpha(p',p) \ =\ \Gamma_\alpha + \frac 
{\Gamma_\alpha (p'^2 + p^2) + p'\!\!\!\!/ \Gamma_\alpha p\!\!\!/}{M^2}\ .
 \ee

\begin{figure}[h]
   \begin{center}
 \includegraphics[width=2.5in]{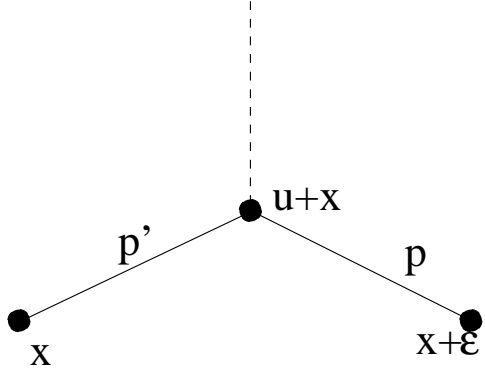}
        \vspace{-2mm}
    \end{center}
\caption{${\mathcal G}_1(x,x+\epsilon)$}
\label{Greenserm}
\end{figure}

  Green's function (\ref{G1}) involves two terms. The first term coming from the structure
$\sim  F_{\alpha\beta} u_\beta$ in the integrand depends only on the difference 
$x - (x+\epsilon)= -\epsilon$. 
The second term  $\sim  F_{\alpha\beta} x_\beta$ depends both on $x$ and $\epsilon$. It is not 
translationally invariant, which is not surprising as 
the fixed point gauge condition breaks translational invariance. For the first term, the calculation gives
  \be
 \label{G1inv}
{\mathcal G}_1^{\rm tr.\, inv.}(-\epsilon)
 \ = \ \int  \frac {d^4p}{(2\pi)^4} G_1(p) e^{ip\epsilon} 
  \ee
with
 \be
\label{G1p}
 G_1(p) \ =\ \frac {F_{\alpha\beta} \Gamma_{\alpha} \Gamma_{\beta } p\!\!\!/ (1 + 2p^2/M^2)}{2p^4(1+ p^2/M^2)^2} 
+ \ldots\ . 
 \ee
We have explicitly written here only the terms involving an irreducible product of
three gamma matrices in  that contribute to the anomaly. The structure 
$\sim F_{\alpha\beta} p_\alpha \Gamma_\beta$ gives zero, being substituted in Eq.(\ref{andir}).

The second translationally noninvariant piece is
 \be
 \label{G1neinv}
{\mathcal G}^{\rm tr.\, noninv.}_1(x, x+\epsilon)
\ =\ - \frac i2 F_{\alpha\beta} x_\beta \int \frac {d^4p}{(2\pi)^4} 
e^{ip\epsilon} \, G_0(p)  \Lambda_\alpha (p,p) G_0(p)\ .
  \ee
 We have
 \be
\label{GLG}
 G_0(p)  \Lambda_\alpha (p,p) G_0(p) \ =\ \frac {\Gamma_\alpha}{p^2(1+ p^2/M^2)} +\ 
{\rm irrelevant\ for\ anomaly\ terms.}
 \ee  
Substituting the sum of (\ref{G1inv}), (\ref{G1neinv}) into Eq.(\ref{andir}) 
and adding to this Eq.(\ref{ii}), we derive 
 \be
 \label{result}
 \partial_\mu {\cal A}_\mu  = - i\lim_{\epsilon \to 0}
      {F_{\mu\nu} \epsilon_\nu} \int \frac {d^4p \, e^{ip\epsilon} }{2(2\pi)^4 p^4}
\left\{ \frac {1+ 2p^2/M^2}{(1 + p^2/M^2)} + \frac {p^2}{M^2(1 + p^2/M^2)} - 
\frac {2p^2}{M^2(1 + p^2/M^2)} \right\}  \nn \\
 \cdot {\rm Tr} \{\Gamma_\mu 
\Gamma_{\alpha} \Gamma_{\beta }  p\!\!\!/ \Gamma^5 \}\, F_{\alpha\beta}  = - \frac 1 {16\pi^2} 
\epsilon_{\mu\nu\alpha\beta}
F_{\mu\nu} F_{\alpha\beta}\ ,
 \ee
the same for all $M$. Note that, for the standard Dirac action in the limit $M \to \infty$, the only
contribution to the anomaly comes from the translationally invariant piece in ${\mathcal G}_1$ of
Eq.(\ref{G1inv}). But when $M < \infty$, all three  contributions discussed above are important. 

Cancellation of $M$ dependence in the sum of the three contributions should work also in higher dimensions. 
Only one has to 
consider higher terms of the expansion of Green's function in $F$. For example, for $D=6$, the relevant 
contributions come from:
 {\it (i)} substituting in  Eq. (\ref{andir}),
 with covariant derivatives
traded for the usual ones, the translationally invariant and {\it (ii)} translationally noninvariant parts of 
${\mathcal G}_2$;
{\it (iii)} substituting in Eq. (\ref{andir}), with the terms $\sim F$ in Eq.(\ref{DFSig}) taken into account, 
the translationally invariant part of ${\mathcal G}_1$,

\end{document}